\documentclass{article}
\usepackage[utf8]{inputenc}
\usepackage{graphicx}
\usepackage{caption}
\usepackage{amsmath,pdftexcmds}
\usepackage[a4paper, total={6.5in, 9.5in}]{geometry}
\usepackage{color}
\usepackage{braket}
\usepackage{comment}
\usepackage{abstract}
\usepackage{amsfonts}
\usepackage{subfig}
\usepackage{graphicx}
\usepackage{multirow}
\usepackage{caption}
\usepackage{pdfpages}
\usepackage{amssymb}
\usepackage{float}
\usepackage{pgfplots}
\usepackage{tikz}
\usepackage{tikz-3dplot}
\usepackage{authblk}
\usepackage[none]{hyphenat}
\usepackage{epsfig}
\graphicspath{graphics/}
\usepackage[colorlinks,linkcolor={black},citecolor={black},urlcolor={blue},unicode,breaklinks=true]{hyperref}

\title{\textbf{Two-states Brownian particle in a Harmonic Potential}}

\author{Giovanni Battista Carollo, Giuseppe Gonnella, Daniela Moretti, Antonio Suma}
\affil{Dipartimento Interateneo di Fisica, Universit\'a degli Studi di Bari and INFN, Sezione di Bari, via Amendola 173, Bari, I-70126, Italy.}
\author{Fulvio Baldovin, Enzo Orlandini}
\affil{Dipartimento di Fisica e Astronomia ‘G. Galilei’ - DFA, Universit\'a di Padova and INFN, Sezione di Padova, Via Marzolo 8, 35131 Padova (PD), Italy}

\pgfplotsset{compat=1.18}
 
\begin{document}

\maketitle

\begin{abstract}
We study the behaviour of a Brownian particle in the overdamped regime in the presence of a harmonic potential, assuming its diffusion coefficient to randomly jump between two distinct values. In particular, we characterize the probability distribution of the particle position and provide detailed expressions for the mean square displacement and the kurtosis. 
We highlight non-Gaussian behavior even within the long-term limit carried over with an excess of probability both in the central part and in the distribution's tails. Moreover, when one of the two diffusion coefficients assumes the value zero, we provide evidence that the probability distribution develops a cusp.
Most of our results are analytical, and corroborated by numerical simulations.
\end{abstract}

\section{Introduction}
\indent Since the seminal work by Einstein \cite{Einstein}, diffusion has become a central concept in physics.
The mathematical frameworks required to address diffusion are by now well established; nonetheless, contemporary research continues to uncover intriguing diffusive processes.
As pointed out in \cite{Chechkin_2017}, a normal free diffusion has two important properties: 1) a Gaussian Probability Density Function (PDF); 2) a Mean Square Displacement (MSD) linearly growing in time (\textit{Brownian} or \textit{Fickian} property). Violating one of these properties leads to \textit{anomalous}  diffusion \cite{Bouchaud1990,Metzler2001,sposini2022robust}. 
In the past, a lot of effort has been paid to studyanomalous diffusion emerging from the violation of the property 2), as reported by various experiments \cite{Banks+Frandin,Barkai_2011,Golding_2006,Metzler_2016}. 
In these cases, the MSD is often proportional to $t^\alpha$, where $\alpha\neq1$ is a real number. 
Important theoretical models capable of reproducing this anomalous time dependence are Lévy flights \cite{Metzler2001}, the Continuous Time Random Walk, and the Fractional Brownian Motion \cite{Barkai}.
From a rigorous point of view, the Central Limit Theorem \cite{Fischer} ensures that the sum of independent and identically distributed random variables possessing a finite second moment has a Gaussian PDF, explaining why it plays a privileged role in the theory of probability. For this reason, stochastic processes violating property 1) received less attention than those violating property 2). 
However, many experiments showed that this effect is real and widespread in different contexts \cite{Wang_pnas_2009,Wang_prl_2014,Hapca_2009,Chubynski,Sposini_2024_1,Sposini_2024_2}.
This type of diffusion is called \textit{Brownian yet non-Gaussian} and it can be modeled allowing the diffusion coefficient $D$ to be a stationary process or to fluctuate in space \cite{Postnikov_2020}. For instance, the so-called diffusing diffusivity models provide examples with linear MSD and non-Gaussian PDF in which $D(t)$ evolves stochastically in time. 
In this framework, some tricks must be imposed on the diffusion coefficient to prevent it from being negative, for example imposing specific boundary conditions, or assuming $D(t)$ to be the square of a diffusion process~\cite{Chechkin_2017,Jain}.
Other models having a diffusion coefficient depending on time have been recently introduced to study the motion of the center of mass of objects where the diffusion coefficient scales with their size, for example, clusters of Brownian particles \cite{Luczka} or polymers in a fluid \cite{Baldovin_2019,Nampoothiri_2021,Nampoothiri_2022,Marcone_2022}. 
In addition to free diffusive systems, confined systems where fluctuations play a crucial role have also received significant attention. 
The development of optical tweezers \cite{Gieseler:21,Volpe_2023}, which use laser light to trap mesoscopic particles in a fluid, has enabled the experimental study of these systems. From a mathematical standpoint, an external potential is introduced in the equation of motion, the simplest and most paradigmatic being the harmonic potential. 
The confinement significantly impacts the behavior of the diffusing particle: the distribution of the position of an overdamped Brownian particle trapped in a harmonic potential, i.e. Ornstein-Uhlenbeck (OU) particle, remains Gaussian, but its variance is not linear in time anymore, approaching instead a constant value at long times \cite{vanKampen}. 
More drastically, a non-harmonic potential usually destroys  Gaussian fluctuations.

\indent In this paper we are interested in the study of an OU particle in a harmonic trap, with the diffusion coefficient $D(t)$ being a stochastic process \cite{Akimoto}. 
To maintain the treatment as simple as possible, we assume $D(t)$ to be sampled from a 2-state telegraph process.
From the experimental point of view, this theoretical setup can be justified by looking at a tracer with a size taking two different values, randomly. Similar processes have been used to describe a protein or a DNA filament switching between a folded and an unfolded configuration \cite{Dieterich}. 
In this context, we consider the Langevin equation as a given evolution equation, irrespective of its physical origin and without reference to the fluctuation-dissipation relation \cite{vanKampen}. This omission means that the mobility of the overdamped particle does not depend on the diffusion coefficient, and consequently,  on time. 
We will show that the PDF of the particle's position $x(t)$ develops non-Gaussian tails. Remarkably, unlike free diffusion, where the PDF tends to a Gaussian distribution in the long-time limit~\cite{Nampoothiri_2022,Sposini_2024_2}, in the confined system it remains always non-Gaussian. 
To quantify this effect, we compute the first central moments, evaluating the deviation from Gaussianity using the kurtosis. 
Finally, as already pointed out in the absence of confinement \cite{Burov,Doerries_1,Doerries_2,Doerries_3,Postnikov_2020}, we highlight that the PDF of the particle location shows a cusp singularity at the origin when the smallest value for the diffusion coefficient goes to 0.
Our theoretical setup has also the advantage of being fully analytically treatable.

\indent The paper is organized as follows. In section \ref{model} we introduce the mathematical description of a harmonic oscillator in the presence of a Brownian motion with a diffusion coefficient sampled from a telegraph process. 
We remark that the position of the particle $x(t)$ can be interpreted as a subordinated process \cite{Feller}, where the subordinator is the stochastic process describing the evolution of the diffusion coefficient. 
Employing this idea, in section \ref{centralmomentsxt} we compute the central moments for the PDF of $x(t)$. Next, in section \ref{probxt} we use our results to study the properties of the PDF of $x(t)$. We conclude the paper with general comments in section \ref{conclusion}.

\section{The model}
\label{model}
Let us consider an OU process \cite{vanKampen,Risken} in one-dimension:
\begin{equation}
    \dot x(t)= -\kappa x(t)+\sqrt {2D(t)} \xi(t)\,,
    \label{xdynamics}
\end{equation}
where $\xi(t)$ is a Gaussian white noise ($\langle \xi(t)\rangle=0$, $\langle \xi(t)\xi(t')\rangle=\delta(t-t')$), and $\kappa>0$. We assume the diffusion coefficient $D(t)$ to evolve in time according to a telegraph process between two nonnegative values $D_1$ and $D_2$, described by the master equation
\begin{equation}
    \dot {\vec{p}}_D(D,t|D_0,t_0)=W\,\vec{p}_D(D,t|D_0,t_0) \ .
    \label{Ddynamics}
\end{equation}
Here, $\vec p_D(D,t|D_0,t_0)=(p_D(D_1,t|D_0,t_0),p_D(D_2,t|D_0,t_0))$ is the vector of the probabilities that $D(t)$ at the instant of time $t$ takes one of the values $D_1$, $D_2$, given that it was $D_0$ at time $t_0$, while $W$ is the matrix of transition rates
\begin{equation}
    W=\begin{pmatrix}
        -\lambda_1&\lambda_2\\
        \lambda_1&-\lambda_2
    \end{pmatrix} \ ,
    \label{Wmatrix}
\end{equation}
being $\lambda_1>0$ the jumping rate (probability per unit time) from state $D_2$ to state $D_1$ and $\lambda_2>0$ and vice versa. The characteristic time is then given by $\tau=(\lambda_1+\lambda_2)^{-1}$.
The probability distribution for $D(t)$ is obtained by solving eq.~\eqref{Ddynamics} and its stationary probability (for $t\gg \tau$) is 
\begin{equation}
    \begin{aligned}
        p_{D_\text{stat}}(D_j)=\frac{\lambda_{2}\delta_{j1}+\lambda_{1}\delta_{j2}} {(\lambda_1+\lambda_2)} \ ,
    \end{aligned}
    \label{probDstaz}
\end{equation}
with a variance 
\begin{equation}
    \sigma^2_D=\frac{\lambda_1\lambda_2}{(\lambda_1+\lambda_2)^2}(D_1-D_2)^2 \ ,
    \label{charvartelegraph}
\end{equation}
respectively.
We now focus on the probability distribution $p_x$ of the position $x(t)$ of the particle. If, momentarily, we address the standard case of a static, deterministic diffusion coefficient  $D(t)=D$, it is well known \cite{vanKampen,Risken} that $p_x$ is a Gaussian: if the particle at the instant $t_0=0$ is at $x_0$ we have the \textit{propagator},
\begin{equation}
    p_x(x,t|x_0)=\frac{1}{\sqrt{2\pi \sigma^2(t)}}e^{-\frac{(x-\mu(t))^2}{2\sigma^2(t)}}\equiv G(x,\sigma^2(t)|x_0) \ ,
    \label{probOU}
\end{equation}
with mean $\mu(t)=x_0e^{-\kappa t}$ and variance $\sigma^2(t)=D\frac{1-e^{-2\kappa t}}{\kappa}$, respectively. Here the parameter $\kappa$ plays the role of the inverse of a characteristic time. 
With $\kappa>0$ the long-time behavior of the MSD of $x(t)$ is not linear in time, but equal to the constant value $D/\kappa$.
Note that to be consistent with what follows we use the uncommon notation in the arguments of the Green function in which the (implicit) time-dependence is via $\sigma(t)$.
Below, we will be interested in deviations of $p_x$ from the Gaussian behavior when $D(t)$ is allowed to fluctuate in time. One way to quantify these deviations is by estimating the first central moments of $p_x$, as it is outlined in the next section.

\section{Moments for a subordinated process}
\label{centralmomentsxt}
According to the previous discussion, when the $D(t)$ is a stochastic process the variance of the Green function becomes a random quantity which we indicate as the \textit{subordinator} $s(t)$ (see below). The variable $x(t)$ is subject to a double source of randomness, and its PDF is given by the so-called \textit{subordination formula} \cite{Feller},
\begin{equation}
    p_x(x,t|x_0)=\int_0^\infty ds' G(x,s'|x_0)p_s(s',t) \ .
    \label{subformula}
\end{equation}
The PDF $p_s(s',t)$ indicates the probability for $s(t)$ of having the value $s'$ at time $t$. 
This formula represents an average of Gaussian distributions $G(x,s'|x_0)$, over different realizations of the subordinator process $s(t)$.
To find the propagator $\overline{G}$ for a generic diffusion coefficient dependent on time, we solve the Fokker-Planck equation associated with eq.~\ref{xdynamics}, given by 
\begin{equation}
    \partial_t\overline{G}(x,t|x_0)=\kappa \partial_x(x \overline{G}(x,t|x_0))+D(t)\partial_x^2 \overline{G}(x,t|x_0) \ ,
    \label{FokkerPlanck}
\end{equation}
imposing as initial condition $\overline{G}(x,0|x_0)=p(x,0)=\delta(x-x_0)$. This equation can be solved in Fourier space, where it reads
\begin{equation}
    \partial_t \overline{G}(k,t|k_0)=-\kappa k\partial_k \overline{G}(k,t|k_0))-D(t)k^2 \overline{G}(k,t|k_0) \ ,
    \label{FokkerPlanckFourier}
\end{equation}
and then using the method of characteristics. The two characteristics equations for the auxiliary parameter $u$  are $\frac{dt}{du}=1$ and $\frac{dk}{du}=\kappa\,k$, with solutions $t=u$ and $k=k_0e^{\kappa u}$, respectively. 
In such a way, eq.~\eqref{FokkerPlanckFourier} becomes an ordinary differential equation for the auxiliary parameter, namely $\frac{d}{du}\overline{G}(u)=-D(u)k_0^2e^{2\kappa u}\overline{G}(u)$. The latter must be solved by imposing the initial condition, reading in Fourier space  $\overline{G}(0)=e^{ik_0x_0}$. Then, the solution is  $\overline{G}(u)=e^{ik_0x_0}e^{-k_0^2 \int_0^{u} du' e^{2\kappa u'}D(u')}$, where $k_0$ and $u$ must be substituted from the characteristic equations. We finally get in Fourier space
\begin{equation}
    \overline{G}(k,t|x_0)=e^{ike^{-\kappa t}x_0}e^{-k^2e^{-2\kappa t} \int_0^{t} du' e^{2\kappa u'}D(u')} \ ,
    \label{propagatorFourier}
\end{equation}
which, after anti-transforming, becomes a Gaussian analogously to eq.~\eqref{probOU}, but with the variance depending on the diffusion coefficient in a more complex way:
\begin{equation}
    s(t)=2e^{-2\kappa t}\int_0^{t}du'e^{2\kappa u'}D(u') \ .
    \label{sprocess}
\end{equation}
Note that, for a deterministic, constant diffusion coefficient,  $s(t)$ reduces to the variance $\sigma^2(t)$ defined in the previous section.
It is important to emphasize that the formula of the propagator does not imply that the probability distribution of $x(t)$ is Gaussian. 
In each realization of the random process $D(t)$, the PDF $G(x,s|x_0)$ is a Gaussian with
a deterministic mean and the variance depending on the particular realization of $D(t)$  according to eq. \eqref{sprocess}. Then eq. \eqref{subformula} corresponds to averaging over realizations,
which results in a non-Gaussian distribution whenever the diffusion coefficient is not a deterministic constant and thus $p_s(s,t)\neq\delta(s-D\frac{1-e^{-2\kappa t}}{\kappa})$.

Using the subordination formula, by imposing the Green function for the subordinated process and performing the change of variables $x\rightarrow x-\mu(t)$, the central moments for $p_x(x,t|x_0)$ are given by 
\begin{equation}
\begin{aligned}
    \mathbb E[(x-\mu(t))^m]=G_{(m)}\mathbb E_D[s^{\frac{m}{2}}(t)] \ , 
\end{aligned}
    \label{centralmoments}
\end{equation}
where $m$ is a positive integer, $\mathbb E_D[\cdot]$ indicates the average over $p_D(t)$, and $G_{(m)}$ is a number defined from the integral $ G_{(m)}=\int_{\mathbb R}dx\frac{e^{-\frac{x^2}{2}}}{\sqrt{2\pi}}x^m $.

\section{Properties of the probability distribution of \textit{x(t)}}
\label{probxt}
\begin{figure}
    \centering
    \includegraphics[width=0.40\textwidth]{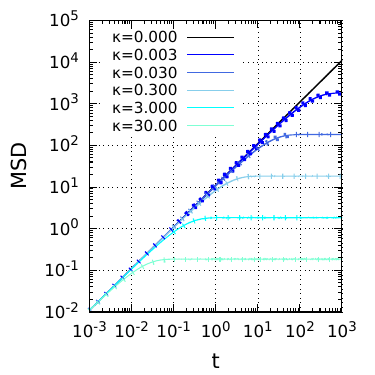}
    \includegraphics[width=0.40\textwidth]{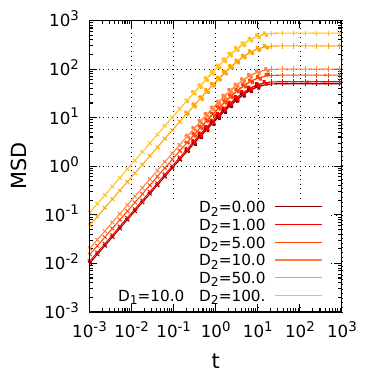}
    \caption{MSD as a function of the time as a function of the relevant parameters change. The continuous lines are the numerical solutions, while the dashed lines are the analytical ones. In the left panel we fixed $D_1=10 D_2$, $\lambda_1=\lambda_2$, and $\tau=10$, and in the right panel we fixed $D_1=10$, $\kappa=0.10$, $\lambda_1=\lambda_2$, and $\tau=10$.}
    \label{fig:1}
\end{figure}

Central moments are basic quantities characterizing the features of the PDF of $x(t)$. Taking advantage of eq. \eqref{centralmoments}, the MSD can be written as
\begin{equation}
\begin{aligned}
    \mathbb E[(x-\mu(t))^2|D_\text{initial}]&=\mathbb E_D[s(t)|D_\text{initial}]=2e^{-2\kappa t}\int_0^{t}ds'e^{2\kappa s'}\mathbb E_D[D(s')|D_\text{initial}]=\\
    &=2e^{-2\kappa t}\sum_{i}D_i\int_0^{t}ds'e^{2\kappa s'}p_D(D_i,s'|D_\text{initial},0) \ ,
\end{aligned}
    \label{MSD}
\end{equation}
where the index $i$ in the sum labels all the possible states of $D(t)$. 
\newline
If at $t_0=0$ $D_\text{initial}$ is sampled from its stationary distribution, eq. \eqref{probDstaz}, the result is
\begin{equation}
    \mathbb E[(x-\mu(t))^2]=\mathbb E_{D_{\text{stat}}}[s(t)]=2e^{-2\kappa t}D_\text{stat}\int_0^{t}ds'e^{2\kappa s'}=D_\text{stat} \frac{1-e^{-2\kappa t}}{\kappa} \ ,
    \label{s1avestaz}
\end{equation}
where $D_\text{stat}=\mathbb E_{D_{\text{stat}}}[D]=\sum_{i}D_ip_\text{stat}(D_i)$. 
In our simple case (two-state process) the stationary value is $D_\text{stat}=\frac{\lambda_1D_2+\lambda_2D_1}{\lambda_1+\lambda_2}$. Notice that in the limit $\kappa\rightarrow 0$ we have $\mathbb E[(x-\mu(t))^2]=2D_\text{stat}t$, linear in time as expected for the free case even in the presence of subordination \cite{Nampoothiri_2022}. The dependence of the MSD on the parameters $\kappa$ and $\tau$, as well as the ratio $D_1/D_2$ is shown in Figure \ref{fig:1}. 
The potential strength $\kappa$ strongly affects the asymptotic value of the MSD. At the same time, the characteristic time $\tau$ is not too relevant once the ratio $\lambda_1/\lambda_2=1/5$ is set: even though the various curves are not the same, varying $\tau$ does not change the short and long time behaviors. It only minimally affects the transient regime. 
At the beginning, all the curves follow a master curve, as $\mathbb E[(x-\mu(t))^2]\sim 2D_\text{stat} t$, which is independent on $\kappa$ and only dependent on the rates and on the values $D_1$ and $D_2$ via their ratio $\lambda_1/\lambda_2$, and $D_1/D_2$, fixed in the panels at the center and on the right.
The non-Gaussianity of  the PDF of $x(t)$ is  typically quantified by deviations from the value $3$ of the kurtosis, 
\begin{equation}    
    K_x(t)=\frac{\mathbb E[(x-\mu(t))^4]}{(\mathbb E[(x-\mu(t))^2])^2}=3\frac{\mathbb E_D[s^2(t)]}{\mathbb E_D[s(t)]^2} \ ,
    \label{kurt}
\end{equation}
having used eq. \eqref{centralmoments}. Similar to the calculation of the MSD, here we have
\begin{equation}
\begin{aligned}
    \mathbb E_D[s^2(t)|D_\text{initial}]&=8e^{-4\kappa t}\int_0^{t}ds''e^{2\kappa s''}\int_0^{s''}ds'e^{2\kappa s'}\mathbb E_D[D(s'')D(s')|D_\text{initial}]\\
    &=8e^{-4\kappa t}\sum_{i,j}D_iD_j\int_0^{t}ds''e^{2\kappa s''}\int_0^{s''}ds'e^{2\kappa s'}p(D_i,s''|D_j,s')p(D_j,s'|D_\text{initial},0)\ ,
\end{aligned}
    \label{s2aveformula}
\end{equation}
where, as before, the indices $i$ and $j$ run over the possible states of $D(t)$. 
\newline
If $D(t)$ is sampled from its stationary distribution as an initial condition, we have
\begin{equation}
    \mathbb E_{D_{\text{stat}}}[s^2(t)]=8e^{-4\kappa t}\sum_{i,j}D_iD_j p_{\text{stat}}(D_j)\int_0^{t}ds''e^{2\kappa s''}\int_0^{s''}ds'e^{2\kappa s'}p(D_i,s''|D_j,s') \ .
    \label{s2avestazformula}
\end{equation}

\begin{figure}
    \centering
    \includegraphics[width=0.40\textwidth]{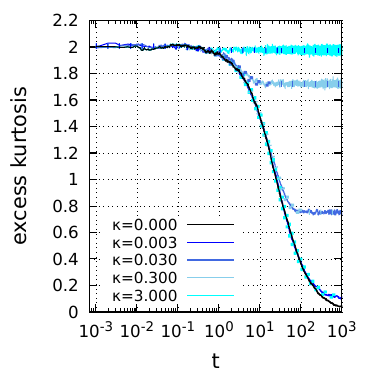}
    \includegraphics[width=0.40\textwidth]{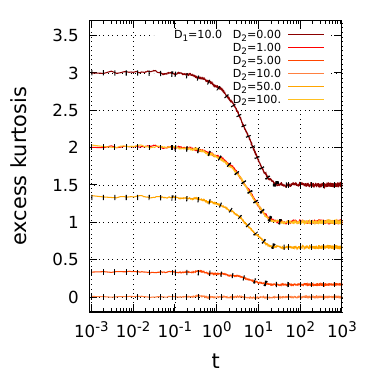}
    \caption{Excess kurtosis as a function of the time as a function of the relevant parameters change. The continuous lines are the numerical solutions, while the dashed lines are the analytical ones obtained inserting eqq. \ref{s1avestaz} and  \ref{s2avestazformula} into eq. \ref{kurt}.
    In the left panel we fixed $D_1=10 D_2$, $\lambda_1=\lambda_2$, and $\tau=10$; and in the right panel we fixed $D_1=10$, $\kappa=0.10$, $\lambda_1=\lambda_2$, and $\tau=10$.}
    \label{fig:2}
\end{figure}

\begin{figure}
    \centering
    \includegraphics[width=0.40\textwidth]{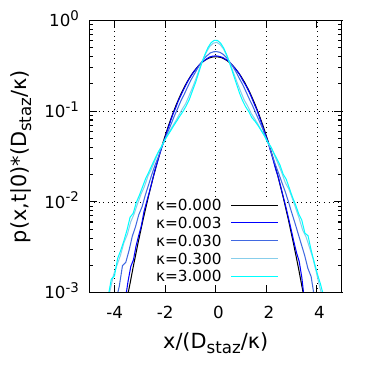}
    \includegraphics[width=0.40\textwidth]{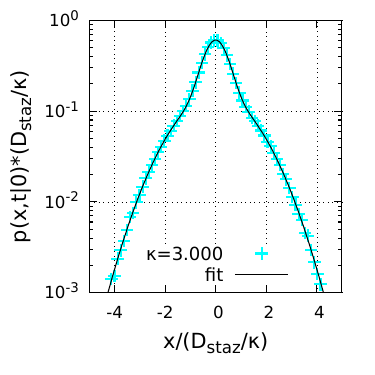}
    \includegraphics[width=0.40\textwidth]{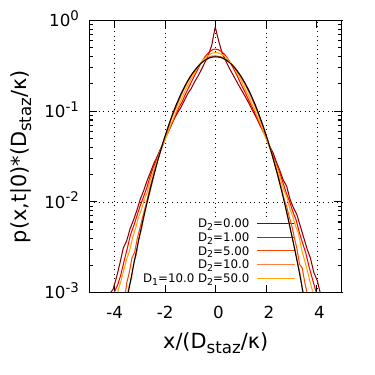}
    \includegraphics[width=0.40\textwidth]{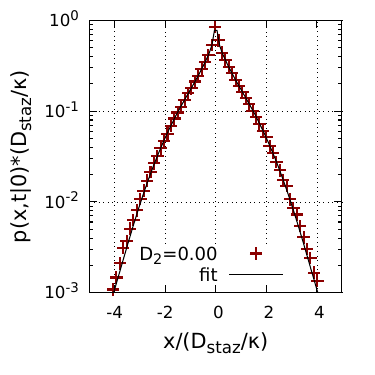}
    \caption{The probability distribution $p_x$ in the stationary regime (t=80$\tau$ on the left, and t=40$\tau$ on the right), obtained using the numerical simulations (the colored ones), with the dashed and the continuous black lines being the Gaussian distributions and the fitting curve, respectively. 
    All curves have been rescaled by the factor $D_{\mathrm{staz}}/\kappa$, being the asymptotic values of the MSD.
    The fit for the cyan-coloured pdf (with $\kappa=3.0$, $D_1=10D_2$, $\lambda_1=\lambda_2$, and $\tau=10$) has been performed as a sum of two Gaussian distributions equally weighted, with variances given by the corresponding asymptotic value of MSD of each state $\{D_1,D_2\}$. 
    Instead, the fit of the brown-coloured pdf (with $\kappa=0.10$, $D_1=10.0$, $D_2=0.0$, and $\lambda_1=\lambda_2$, and $\tau=5$) has a more complex behaviour, as a cusp is present at $x=0$, resulting in a sum of Gaussian, Laplace, and highly-narrow distributions, as shown in 
    \cite{Burov,Postnikov_2020}. }
    \label{fig:3}
\end{figure}

The behavior of the excess kurtosis $K_x(t)-3$ starting from a stationary distribution for $D(t)$ is reported in Figure \ref{fig:2}. 
The analytical curves have been computed by inserting eq. \eqref{s1avestaz}, and eq. \eqref{s2avestazformula} in eq \eqref{kurt}.
We see that the kurtosis tends to a fixed value in the long time limit, which depends on the rates $\lambda_1$, $\lambda_2$, on $D_1$ and $D_2$, and the potential strength $\kappa$:
\begin{equation}
K_x(t\rightarrow\infty)=\frac{3 (\lambda_1+\lambda_2) \left(D_1^2 \lambda_2 (\lambda_2+2 \kappa)+2 D_1 D_2 \lambda_1 \lambda_2+D_2^2 \lambda_1 (\lambda_1+2 \kappa)\right)}{(\lambda_1+\lambda_2+2 \kappa) (D_1 \lambda_2+D_2 \lambda_1)^2} \ ,
\label{asymptkurt}
\end{equation}
independently of the initial distribution of $D(t)$. This means that the distribution of $x(t)$ always remains non-Gaussian during the whole evolution. 
It is easy to check that this expression simplifies to 3 (Gaussian behavior) either when $\kappa=0$, i.e. in the free case, or $D_1=D_2$, i.e. when there is no subordination.
Notice also that the distribution is always leptokurtic, namely its tails decay slower than a Gaussian, even in the long-time limit. This is immediately visible by looking at the distribution of displacements, in Figure \ref{fig:3}. 
As $\kappa$ grows, the distribution develops slowly decaying tails, implying a positive excess kurtosis. 
As claimed in Ref.\cite{Burov}, these tails have a Gaussian behavior (they decay as $e^{-x^2}$, but maintain a probability excess with respect to a Gaussian with the same width). 
Also at $x=0$ the distributions show a probability excess with respect to  Gaussianity. Moreover, when a diffusion coefficient reaches the value 0, the distribution displays a cusp at $x=0$, similar to what is proved in Reff. \cite{Burov,Postnikov_2020} for $\kappa=0$.
These features can be understood on analytical grounds. Indeed, the excess kurtosis is always non-negative \cite{Sposini_2024_2,Sposini_2024_1} irrespective of the distribution chosen for $D(t)$, since $    K_x(t)-3 = 3\frac{\mathbb E_D[s^2(t)]-\mathbb E_D[s(t)]^2}{\mathbb E_D[s(t)]^2} = 3\frac{\mathbb E_D[(s(t)-\mathbb E_D[s(t)])^2]}{\mathbb E_D[s(t)]^2}\ge 0 $.
A possible way to have this expression equal to zero is to impose $s(t)$ to be a deterministic quantity, at least after a certain instant during the time evolution. 
For example, having in mind a discrete Markovian evolution for $D(t)$ this is possible if there are absorbing stationary states. In our case, setting $\lambda_1=0$ or $\lambda_2=0$ implies that once $D(t)$ reaches a given (absorbing) state, it is impossible to jump into the other. 
It is easy to check that eq. \eqref{asymptkurt} follows this rule, namely, setting at least one rate to 0 then $K_x(t\rightarrow\infty)=3$, so the PDF of $x(t)$ must be Gaussian in the long-time limit. Finally, using the Jensen inequality \cite{Sposini_2024_2} one can show that the probability distribution around $x(t)=0$ is always bigger than a Gaussian PDF.

\begin{figure}
    \centering
    \includegraphics[width=0.40\textwidth]{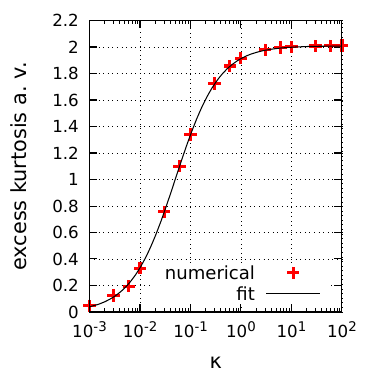}
    \includegraphics[width=0.40\textwidth]{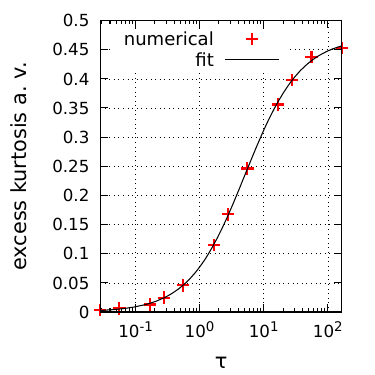}
    \caption{The excess kurtosis $K_x(t)-3$ in the long-time limit as function of $\kappa$ and $\tau$, for parameters equal to the ones of fig. \ref{fig:2} on the left and at the center, obtained from the same numerical simulations. The fit has been performed using a function $f(x)=\frac{ax}{1+bx}$ in the parameters $a$ and $b$, the functional shape derived at eq. \eqref{asymptkurt1}.}
    \label{fig:4}
\end{figure}

The expression of the final value of the kurtosis, eq. \eqref{asymptkurt}, can be simplified by substituting $\tau$ and $\sigma_D$ in place of $\lambda_{1/2}$ giving
\begin{equation}
    K_x(t\rightarrow\infty)-3=\frac{6 \sigma_D^2}{D^2_\text{stat}}\cdot\frac{\kappa \tau}{1+\kappa \tau} \ .
    \label{asymptkurt1}
\end{equation}
This result nicely links the final value of the kurtosis exclusively to the properties of the subordinator process and the potential strength $\kappa$. 
In figure \ref{fig:4} we present a verification for this result.

\section{Conclusions}
\label{conclusion}
\indent Anomalous diffusion has received great attention in recent decades and a new class of models was studied by letting the diffusion coefficient fluctuate in time. In the absence of spatial confinement, one obtains a process having an MSD which is linear in time, but with a non-Gaussian PDF in the small regime.
In this context, we have studied the effect of a harmonic potential in the dynamics, choosing a telegraph process to describe the evolution of the diffusion coefficient. 
This amounts to considering an OU process whose diffusion coefficient fluctuates in time between two values. We have shown that this confinement drastically affects the dynamics compared to the free case. The MSD of the process tends to a constant value, something expected and also present in the case of $D$ deterministically fixed, and the PDF of the particle's position not only becomes non-Gaussian, but, more surprisingly, this non-Gaussianity is stabilized in time.
Indeed, unlike the free case, where the PDF tends to return Gaussian in the long-time regime, the PDF remains always non-Gaussian. 
We envisage that this time persistence would help measure non-Gaussian properties in experimental protocols that usually cannot access short-time behaviours where deviations of Gaussianity normally reside.
In addition, we have shown that if the process alternates diffusive and non-diffusive regimes (i.e. one of the diffusion coefficients becomes 0), the PDF of the particle's position displays a cusp at $x=0$. 

Note that all the findings presented have been obtained in one dimension, as the generalization to higher dimensions is straightforward and mostly results in multiplicative factors.

Finally, we remark that in situations where the fluctuation-dissipation theorem holds and the particle's mobility depends on $D(t)$, the resulting double subordination process would require a different mathematical approach.

\section*{Acknowledgements}
Giovanni Battista Carollo is financially supported by Apulia Region via the initiative \textit{Dottorati di ricerca in Puglia XXXVII ciclo}.

\end{document}